\documentclass[12pt,a4paper]{article}
\usepackage[T1,T2A]{fontenc}
\usepackage[utf8]{inputenc}
\usepackage[russian]{babel}
\usepackage{indentfirst}
\usepackage[a4paper,top=2cm,bottom=2cm,left=3cm,right=2cm,marginparwidth=1.75cm]{geometry}
\usepackage{titlesec}
\titlelabel{\thetitle.\,\,}
\usepackage{amsmath}
\usepackage{amssymb}
\usepackage{graphicx}
\usepackage{color}
\usepackage{enumitem}
\usepackage{tabularx}

\title{\textbf{Синхронизация и десинхронизация в ансамблях мобильных агентов}}
\author{Е.М. Варварин, Г.В. Осипов}
\date{Декабрь 2024}

\begin{document}

\maketitle

\setlength{\parindent}{2.5cm} \setlength{\hangindent}{2cm} \textbf{Аннотация.} 
Анализируются механизмы возникновения и разрушения, а также характеристики синхронного и асинхронного режимов поведения ансамблей (роев) взаимодействующих мобильных агентов, двигающихся согласно хаотическим фазовым траекториям систем Ресслера и Лоренца. Продемонстрирована основанная на эффекте хаотической фазовой синхронизации возможность организации последовательного и параллельного движения агентов.

\setlength{\parindent}{2.5cm} \setlength{\hangindent}{2cm} \textbf{Ключевые слова}: мобильный агент, ансамбль, хаотическая фазовая синхронизация, система Ресслера, система Лоренца.

\section{Введение}

\setlength{\parindent}{0.6cm}

В последние годы понятие "мобильный агент" \cite{leader_agent} получило широкое распространение при моделировании процессов, в которых важно взаимное расположение отдельных агентов (частиц) в фазовом пространстве. Рассматриваются популяции мобильных агентов с импульсной связью с целью применения в системах роевой робототехники и мобильных сенсорных сетях \cite{pulse_coup}. Развивается и само понятие мобильного агента, разработана гибридная модель многоагентных систем, объединяющая преимущества метрических и топологических правил взаимодействия \cite{topolog_metric_coup}. Условия включения связи между агентами также могут быть совершенно разнообразны, помимо взаимодействия на определенном расстоянии \cite{radius_coup, radius_coup_2}, элементы могут взаимодействовать только с теми, кто входит в их поле зрения \cite{cone_coup} или же при нахождении в заранее определенных зонах пространства \cite{zone_coup}. В данной работе на примере хаотических аттракторов Ресслера и Лоренца демонстрируется возможность организации последовательного и параллельного движения агентов на основе эффекта хаотической фазовой синхронизации.
%


\section{Модель}
В качестве мобильного агента рассмотрим материальную точку,
движущуюся в трехмерном пространстве $(x, y, z)$ так,
что ее траектория полностью совпадает с траекторией поставленного ей в соответствие осциллятора. Обобщенный вид, описывающий поведение ансамбля взаимодействующих частиц следующий:

 \begin{equation}
 \begin{cases}
 \dot{x}_i = f_x + d_x \big[ \sum_{j=1}^N (x_j-x_i) \big]\\
 \dot{y}_i = f_y + d_y \big[ \sum_{j=1}^N (y_j-y_i) \big],\qquad \qquad i=\overline{1,N}\\
 \dot{z}_i = f_z + d_z \big[ \sum_{j=1}^N (z_j-z_i) \big]\\
 \end{cases}
 \label{main_eq}
 \end{equation}

В отсутствие связей ($d_x=d_y=d_z=0$) динамика отдельного агента может быть как регулярной, так и хаотической. В общем случае все агенты неидентичные. 

Связь между $i$ - тым и $j$ - тым агентами включается только при их достаточной близости: тогда, когда агенты оказываются внутри цилиндра радиуса $r$:

\begin{equation}
 d_k = 
 \begin{cases}
 d', \: (x_i - x_j)^2 + (y_i - y_j)^2 < r^2 \\
 0, \: \text{в противном случае}
 \end{cases}
 \label{cylinder}
\end{equation}
где $k = \{x,y,z\}$, $d'=const$ -- параметр, определяющий силу связи. Связь между элементами во времени может быть организована тремя способами: 
\begin{enumerate}
\renewcommand{\labelenumi}{\alph{enumi})}
    \item после попадания траекторий в цилиндр связь между осцилляторами не отключается,
    \item связь может быть отключена после как какой-то из осцилляторов покинул цилиндр и
    \item связь может действовать только определенное время.
\end{enumerate}
Если не оговорено специально, в работе рассматривается первый вариант. \\

Целью работы является исследование коллективной динамики ансамблей взаимодействующих агентов, двигающихся по хаотическим траекториям. Рассматривается коллективная динамика двух типов: когерентная синхронная динамика и не когерентная, полностью асинхронная динамика. При когерентной динамике в ансамблях возникает режим хаотической фазовой синхронизации. 
Для режим хаотической фазовой синхронизации двух осцилляторов имеет место выполнении двух условий: совпадение средних частот:

\begin{equation}
 \Omega_i = \langle \nu_i \rangle = \Omega_j = \langle \nu_j \rangle
 \label{omega}
\end{equation}

и наличие ограниченной разности фаз:

\begin{equation}
 |\phi_i(t)-\phi_j(t)| \leq const
 \label{phi_condition}
\end{equation}

В ансамблях синхронизация бывает как глобальной, так и кластерной. В первом случае все агенты двигаются по близким траекториям с некоторым интервалом между друг другом. Это обеспечивается наличием фазового сдвига $\phi_i(t)-\phi_j(t)$. Во втором случае ансамбли разбиваются на отдельные группы синхронно двигающихся агентов. Как будет показано ниже, объединение агентов в синхронные группы может происходить как при притягивающих, так и при отталкивающих связях. В первом случае движение агентов можно рассматривать как последовательной, а во втором как параллельное. Очевидно, что выбирая соответствующим образом топологию связей можно добиться комбинации последовательно и параллельного движения агентов. 

\begin{figure}[h]
 \includegraphics[width=0.18\textwidth]{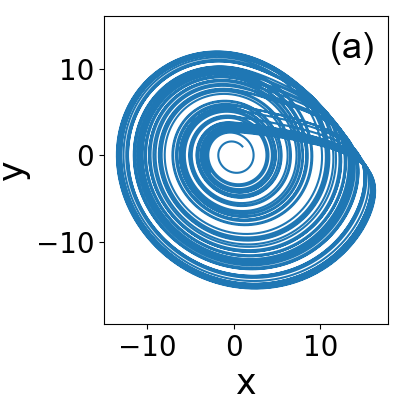}
 \includegraphics[width=0.18\textwidth]{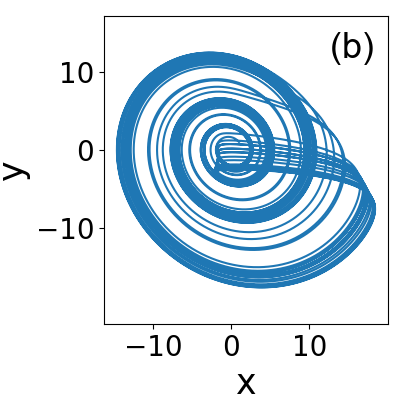}
 \includegraphics[width=0.18\textwidth]{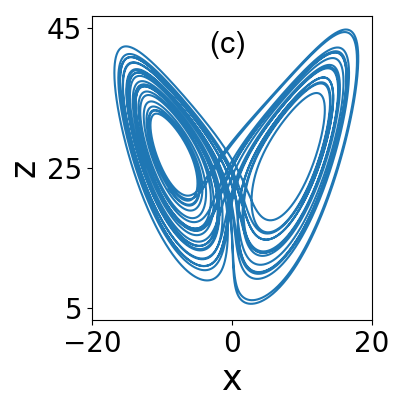}
 \includegraphics[width=0.18\textwidth]{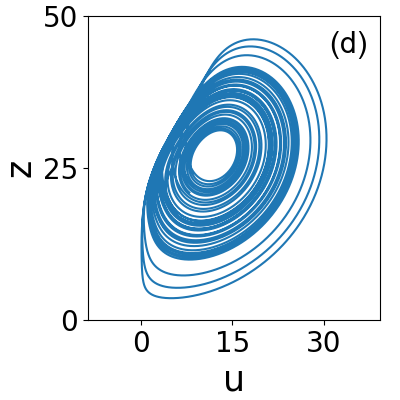}
 \includegraphics[width=0.18\textwidth]{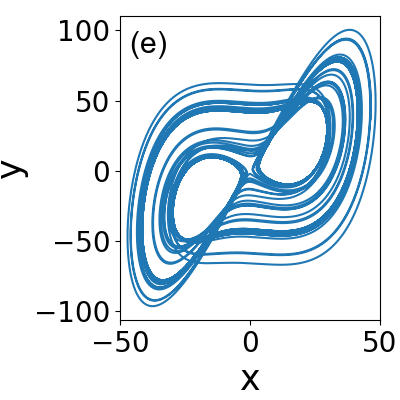}
 \centering
 \caption{\label{Рис. } Проекции хаотических аттракторов. (a) проекция на плоскость $(x,y)$ аттрактора Ресслера (система (\ref{rossler})) в фазо-когерентном режиме ($a=0.16$, $b=0.1$, $c=8.5$), (b) проекция на плоскость $(x,y)$ аттрактора Ресслера в режиме воронка ($a=0.22$, $b=0.1$, $c=8.5$), (c) проекция на плоскость $(x,z)$ классического аттрактора Лоренца (система (12) при $d_y=0$) ($\sigma=10$, $b=8/3$, $r=28$), (d) проекция классического аттрактора Лоренца на плоскость $(u=\sqrt{x^2+y^2}, z)$, (e) проекция на плоскость $(x,y)$ аттрактора Лоренца, возникающий через перемежаемость $\sigma=10$, $b=8/3$, $r=166.1$ }
 \label{attractors_fig}
\end{figure}

В качестве осцилляторов, по траекториям которых двигаются агенты, рассматриваются система Ресслера и система Лоренца. 
Напомним, что в системе Ресслера хаотические колебания возникают в результате каскада бифуркаций удвоения периода предельных циклов. В зависимости от параметров системы хаотический аттрактор бывает фазо-когерентным (phase-coherent) (Рис. \ref{attractors_fig}a) или аттрактором-воронкой (funnel) (Рис.\ref{attractors_fig}b). В системе Лоренца мы рассматриваем классический аттрактор Лоренца (Рис.\ref{attractors_fig}c,d) и хаотический аттрактор, возникающий через перемежаемость (Рис.\ref{attractors_fig}e). 


\section{Синхронизация мобильных агентов.} 
Рассмотрим варианты синхронного поведения ансамблей агентов, движение которых происходит в соответствии с хаотическими траекториями системы Ресслера и системы Лоренца.

\subsection{Система Ресслера}

Пусть движение агента происходит по хаотической траектории системы Ресслера \cite{levin_osipov}:

\begin{equation}
 \begin{cases}
 \dot{x}_i = - w_i y_i - z_i = f_x\\
 \dot{y}_i = w_i x_i + a y_i = f_y,\quad i=\overline{1,N}\\
 \dot{z}_i = b + z_i (x_i - c) = f_z\\
 \end{cases}
 \label{rossler}
\end{equation}
где $a, b, c$ -- положительные параметры. В последующих экспериментах примем $a = 0.16$ 
 для фазо-когерентного аттрактора и $a= 0.22$ для аттрактора воронки, $b = 0.1, c = 8.5$. Параметр $w_i$, выбираемый случайно из интервала $[0.93; 1.07]$ характеризует временные масштабы осцилляций. В нашем исследовании примем: $d' = 0.3, r=4$. Значение параметр связи $d'$ выбрано таким, что все взаимодействующие хаотические осцилляторы синхронизуются по фазе без введения параметра близости траекторий $r$ \cite{osipov2003, pikovsky}.

Параметры системы
выбраны так, чтобы при достаточно близких к нулю
начальных условиях (в настоящей работе рассматривался куб с длиной ребра 10 с центром в начале координат) фазовые траектории не уходили на бесконечность,
а притягивались к хаотическому квазиаттрактору (см, например, \cite{stankevich}). Формально моделирование и указанный
бифуркационный сценарий дают только хаотическое
множество.\\




Введём фазу следующим образом:

\begin{equation}
 \phi = arctan\frac{\dot{y}}{\dot{x}}
 \label{rossler_phi}
\end{equation}

Тогда формула для вычисления средней частоты примет вид:

\begin{equation}
 \langle \dot{\phi} \rangle = \langle \nu \rangle = \langle \frac{\dot{y}\ddot{x}-\ddot{y}\dot{x}}{\dot{x}^2+\dot{y}^2} \rangle
 \label{rossler_omega}
\end{equation}

Было проанализировано влияние параметра $a$ на среднюю частоту системы (Рис. \ref{omega_a_fig}):

\begin{figure}[h]
 \includegraphics[width=0.95\textwidth]{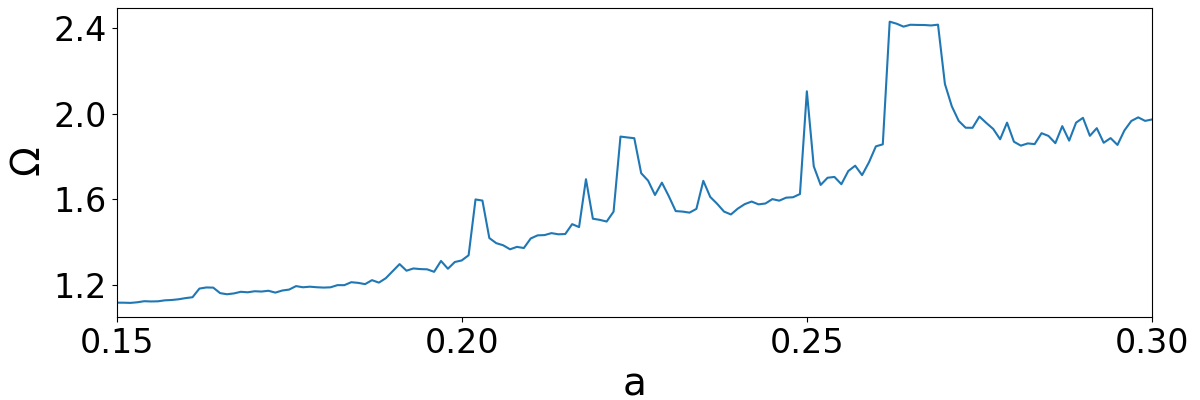}
 \centering
 \caption{\label{Рис. } Зависимость средней частоты от параметра $a$.}
 \label{omega_a_fig}
\end{figure}

Прекрасно видно, что с ростом параметра $a$ средняя частота агента растет. Соответственно, при больших значениях параметра $a$ (при которых аттрактор всё ещё остается хаотическим), скорость синхронизации ансамбля агентов будет значительно выше, чем в случае фазо-когерентного аттрактора ($a=0.16$).

\subsubsection{Связь по переменной $y$}

Рассмотрим коллективное поведения ансамбля из 10 мобильных агентов, движение которых подчиняется хаотической траектории системы Ресслера -- фазокогерентному аттрактору.
Пусть системы Ресслера связаны только по переменной $y$ во втором уравнении системы (\ref{main_eq}), для этого $d_y$ задаётся выражением (\ref{cylinder}), $d_x = 0$, $d_z = 0$.

\begin{figure}[h]
 \includegraphics[width=0.32\textwidth]{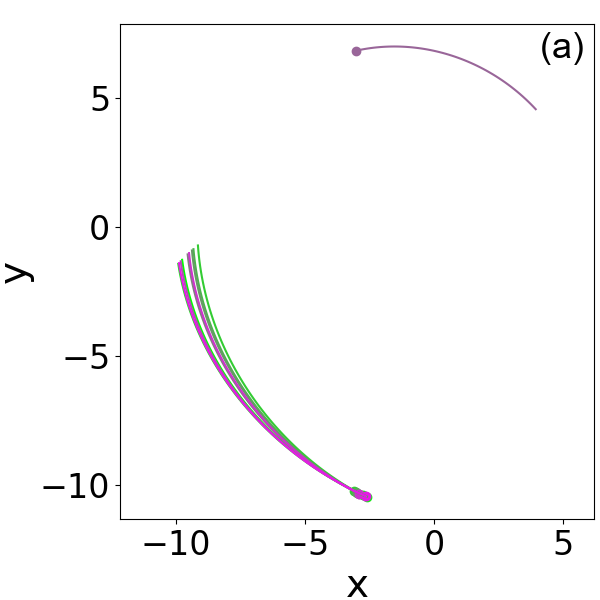}
 \includegraphics[width=0.32\textwidth]{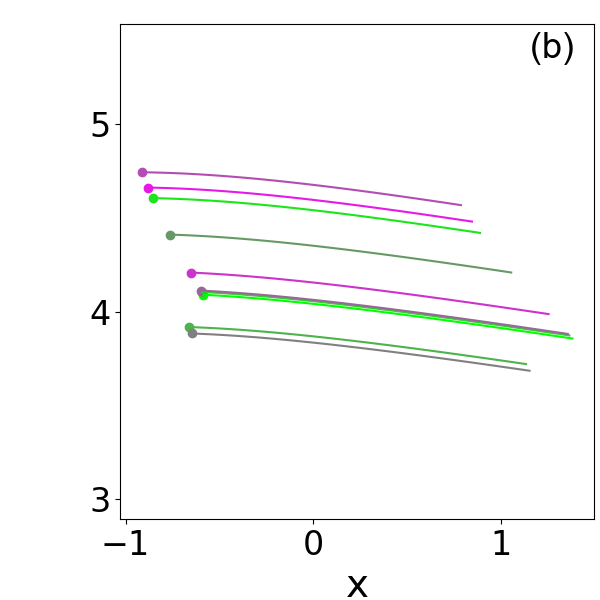}
 \includegraphics[width=0.32\textwidth]
 {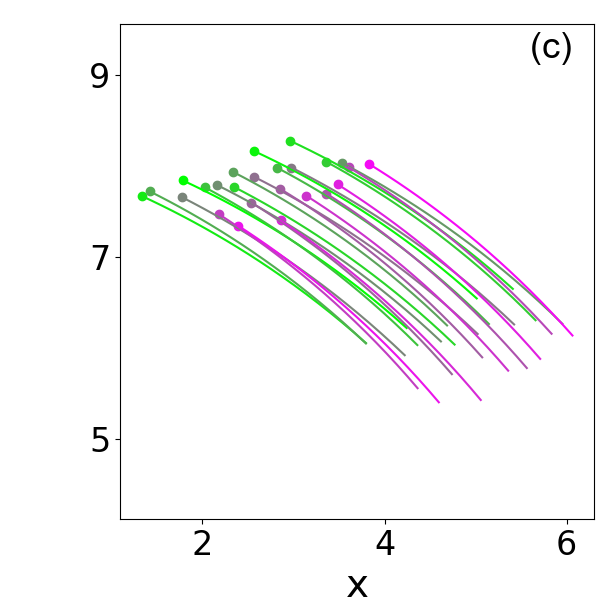}
 \centering
 \caption{\label{Рис. } Синхронизация ансамбля мобильных агентов. (а) последовательное движение агентов (друг за другом), связь по переменной $y$, $d_x=d_z=0$ в системе (\ref{main_eq}) , (b) параллельное движение агентов (единым фронтом), связь по переменной $x$, $d_y=d_z=0$ в системе (\ref{main_eq}) (c) - движение в заданной конфигурации "решетка", скомбинированные связи, аналогичные случаю "Придание рою мобильных агентов структуры различных геометрических форм" в \cite{pjtf}.}
 \label{couplings_figs}
\end{figure}

Первые работы по хаотической фазовой синхронизации связанных систем Ресслера были посвящены анализу систем связанных по переменной $y$. Было обнаружено, что пара \cite{pikovsky} и цепочка \cite{osipovPRE} связанных неидентичных систем Ресслера в хаотическом режиме переходят в синхронный режим при достижении некоторого критического значения параметра связи. В нашем случае связь между элементами включается только тогда, когда фазовые траектории какой-либо пары осцилляторов попадают внутрь цилиндра (\ref{cylinder}). Так как связь выбрана достаточно большой, то осцилляторы синхронизуются. У них имеет место совпадение средних частот и ограниченность фазовой расстройки $|\phi_i(t)-\phi_j(t)| \leq const.$ При этом соответствующие мобильные агенты двигаются по близким траекториям на некотором расстоянии друг от друга (Рис.\ref{couplings_figs}а). (см также \cite{pjtf}.) 

\subsubsection{Связь по переменной $x$}

 Пусть системы Ресслера связаны только по переменной $x$ в первом уравнении системы (\ref{main_eq}), для этого $d_x$ задаётся выражением (\ref{cylinder}), $d_y = 0$, $d_z = 0$. 
Аналогично связи по $y$, с течением времени агенты начнут объединяться, образуя кластеры. В итоге будет достигнута глобальная синхронизация. Но в данном случае движение будет не "по цепочке", а все вместе, "единым фронтом". Поведение агентов похоже на случай "параллельное движение" \cite{pjtf}, однако из-за отсутствия противоположно направленных связей агенты могут приближаться друг с другом на любое расстояние. Результаты представлены динамики ансамбля из 10 элементов на Рис. \ref{couplings_figs}b. 

\subsubsection{Комбинированная связь по переменным $x$ и $y$}

Выше показано, что связь по переменной $y$ обеспечивает последовательной движение агентов, а связь по переменной $x$ обеспечивает параллельное движение. Подбирая соответствующим образом межэлементные связи легко получить всевозможные топологические структуры из синхронизованных агентов. На Рис.\ref{couplings_figs}c представлена конфигурация типа "двумерная решетка" размером 5х5.

\subsection{Синхронизация аттракторов Лоренца}

Используя связи, описанные выше, попробуем применить те же подходы для синхронизации другого хаотического аттрактора -- аттрактора Лоренца, записанного в следующем виде:

\begin{equation}
 \begin{cases}
 \dot{x}_i = \sigma(y_i-x_i)\\
 \dot{y}_i = r_ix_i - y_i - x_i z_i + d_y \cdot \big[ \sum_{j=1}^N(y_j-y_i) \big], \qquad i=\overline{1,N}\\
 \dot{z}_i = -bz_i+x_iy_i
 \end{cases}
 \label{lorenz}
\end{equation}

где $b=\frac{8}{3}$, $\sigma=10$, значения параметра $r$ будут определяться типом аттрактора Лоренца.

Рассмотрим аттракторы Лоренца двух видов:

\begin{itemize}
 \item Классический аттрактор Лоренца. Для системы из пяти элементов возьмем $r_i \in [28; 28.1]$. 
 Формула для фазы известна и вычисляются следующим образом:
 \begin{equation}
 \phi = arctan\frac{z-z_0}{u-u_0},
 \end{equation}
 где $u=\sqrt{x^2+y^2}$, $u_0=12$ и $z_0=27$.
 \item Аттрактор Лоренца, возникающий через перемежаемость (intermittent). В этом случае $r_i \in [166.1; 166.2]$
 \begin{equation}
 \phi(t) = 2\pi \frac{t-t_n}{t_{n+1}-t_n}+2\pi n, t_n \leq t \leq t_{n+1},
 \end{equation}
 Где $[t_n; t_{n+1}]$ -- цикл, состоящий из одной ламинарной и одной турбулентной стадий под номером $n$. Подробнее этот подход описан в \cite{osipov2007}.

\end{itemize}

\begin{figure}[h]
 \includegraphics[height=4cm]{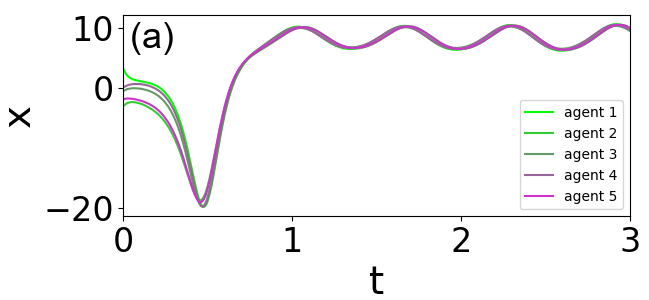}
 \includegraphics[height=4cm]
 {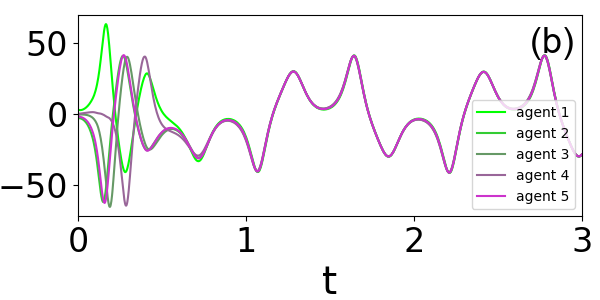}
 
 \caption{\label{Рис. } Синхронизация систем Лоренца. (a) -- синхронизация в случае существования классического аттрактора, (b) -- синхронизация в случае аттрактора, возникшего через перемежаемость. Начальные условия для каждого агента генерировались случайно, $x \in [-5; 5], y\in [-5; 5], z\in [-1; 1]$.}
 \label{lorenz_sync_fig}
\end{figure}

Результаты численного моделирования процесса синхронизации для систем Лоренца, демонстрирующих оба типа хаотического поведения, представлены на Рис. \ref{lorenz_sync_fig}. В обоих случаях, 
как для связанных осцилляторов Ресслера, сначала наступает кластерная синхронизация, а затем глобальная. Связь подобрана таким образом, что имеет место последовательное движение мобильных агентов. Следует отметить, что переход к режиму глобальной синхронизации происходит в течение короткого времени.

\section{Десинхронизация роя}

\subsection{Введение связи по переменной $z$}

В предыдущих примерах стояла задача достижения синхронного режима поведения мобильных агентов. В данном разделе будут обсуждаться способы десинхронизации -- разрушения синхронного режима. Оказалось, что добиться этого можно введением межэелементной связи по переменной $z$ в третье уравнение системы (\ref{main_eq}).

\begin{figure}[h]
 \includegraphics[width=0.99\textwidth]{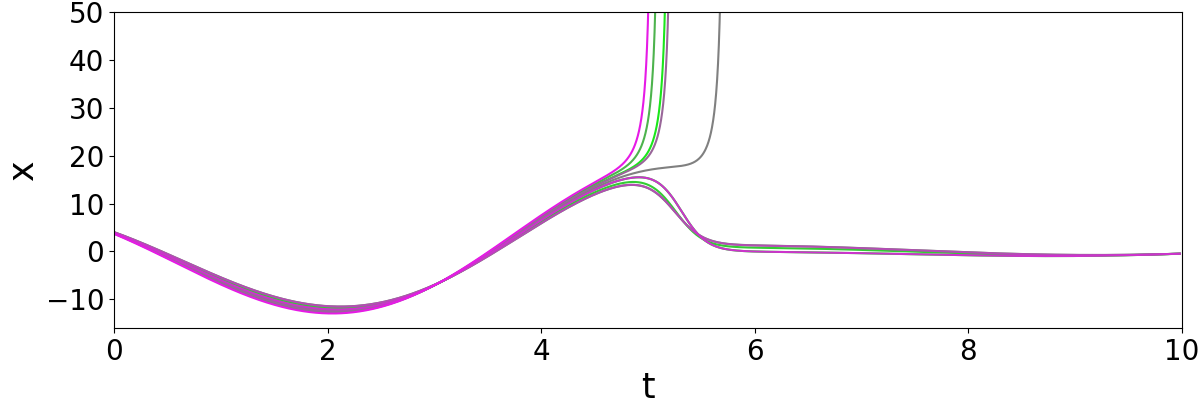}
 
 \caption{\label{Рис. } Добавление в глобально синхронизированную систему связи по $z$. Изначально в аттракторе 10 агентов. Довольно быстро 8 агентов оказываются за пределами области $L$. При $t>6$ синхронизованными остаются только два агента. Через некоторое время (на рисунке момент не указан) между оставшимися агентами происходит десинхронизация.}
 \label{desync_fig}
\end{figure}

 Для демонстрации эффекта десинхронизации был проведен следующий эксперимент: сначала за счет связи по переменной $y$ был обеспечена глобальная синхронизация агентов. Они последовательно двигались (см Рис. 2 (а)) в некоторой локальной движущейся в соответствии с фазовыми траекториями осцилляторов области $L$ фазового пространства. Далее, после включения связи по переменной $z$, агенты в следствии взаимодействия поочередно вылетали за пределы области $L$. Связь по переменной $z$ в такой постановке является отталкивающей. Результаты описанного выше эксперимента представлены на Рис. \ref{desync_fig}.

\subsection{Изменение величины межэлементных связей и уменьшение радиуса действия связи}

Для решения задачи десинхронизации роя возможно применение двух простых алгоритмов. В первом случае разрушить синхронный режим можно уменьшив величину межэлементной связи $d^{'}$. В силу разных положений в момент уменьшения (или отключения) связи хаотические системы не будут синхронизованы: будет иметь место разбегание фазовых траекторий. Во втором случае к десинхронизации может привести уменьшение величины радиуса цилиндра, в котором включена связь между агентами ($r$). Вследствие хаотического поведения мобильных агентов при уменьшении величины $r$ какие-либо хаотические траектории могут покинуть область $L$ и количество синхронизованных элементов сократиться. 

\section{Результаты}

В результате исследования синхронизации и десинхронизации роя мобильных агентов получено следующее:
\begin{itemize}
 \item На примере ансамблей взаимодействующих мобильных агентов, траектории движения которых подчиняются системам Ресслера и Лоренца в хаотических режимах, было рассмотрено влияние различных видов связей на коллективную динамику, в частности на возникновение синхронных режимов. Показана возможность организации последовательного и параллельного движения агентов и организации различных топологических конфигураций роя агентов;
 \item Для случая хаотических аттракторов Ресслера были предложены и успешно протестированы три способа десинхронизации роя: при введение дополнительной связи по переменной $z$, при уменьшении межэелементной связи и при уменьшении радиуса действия связи.
\end{itemize}

\section{Благодарности}

Работа выполнена при финансовой поддержке гранта РНФ \#23-12-00180 (задача синхронизации) и проекта № 0729-2020-0036 Министерства науки и высшего образования Российской Федерации (задача десинхронизации)

\newpage

\end{document}